\documentclass[10pt,journal]{IEEEtran}
\usepackage{cite}
\usepackage{graphicx}
\DeclareGraphicsExtensions{.pdf,.jpeg,.png}
\usepackage[cmex10]{amsmath}
\usepackage{array}
\usepackage{mdwmath}
\usepackage{mdwtab}
\usepackage{fixltx2e}

\begin{document}

  \title{Matrix Filters for the Detection of Extragalactic Point Sources
    in Cosmic Microwave Background Images}

  \author{Diego~Herranz\IEEEmembership{}
    and~Jos\'e~Luis~Sanz~\IEEEmembership{}%
    \thanks{D. Herranz and
    J.L.~Sanz are with the Astronomy Department, Instituto de F\'\i
    sica de Cantabria, CSIC-UC, Av. los Castros s/n, 39005 Santander,
    Spain, e-mail: herranz@ifca.unican.es. J.L.~Sanz is currently on a
    sabbatical year at the ISTI (CNR), Pisa, Italy.}%
    \thanks{Manuscript received ---; revised ---.}}

\markboth{Journal of Selected Topics in Signal Processing, Vol.~xx,
  No.~x, Month Year}%
  {Herranz & Sanz: Matrix Filters for the Detection of Extragalactic
  Point Sources in Cosmic Microwave Background Maps}

\maketitle

\begin{abstract}
In this paper we introduce a new linear filtering technique, the
so-called matrix filters, that maximizes the signal-to-interference
ratio of compact sources of unknown intensity embedded in a set of
images by taking into account the cross-correlations between the
different channels. By construction, the new filtering technique
outperforms (or at least equals) the standard matched filter applied
on individual images. An immediate application is the detection of
extragalactic point sources in Cosmic Microwave Background images
obtained at different wavelengths. We test the new technique in two
simulated cases: a simple two-channel case with ideal correlated color
noise and more realistic simulations of the sky as it will be observed
by the LFI instrument of the upcoming ESA's Planck mission. In both
cases we observe an improvement with respect to the standard matched
filter in terms of signal-to-noise interference, number of detections
and number of false alarms.
\end{abstract}


\IEEEpeerreviewmaketitle

\section{Introduction}

\IEEEPARstart{T}{he} detection of faint pointlike sources is a task
that is common to many branches of Astronomy, from the search for
protostars in gas-rich nebulae to the study of active galactic nuclei
in the confines of the observable universe. Since the angular size of
these objects is smaller than the angular resolution of the telescopes
that are used to observe them, they appear as \emph{point sources}
with the shape of the telescope point spread function.

A case of particular interest is the detection of extragalactic point
sources (EPS) in microwave wavelengths. In the microwave range of the
electromagnetic spectrum, the sky is dominated by the so-called Cosmic
Microwave Background (CMB) radiation, a relic of the hot and dense
first moments of the universe. The study of the CMB is one of the
hottest research topics in modern Cosmology. For a short review on the
CMB, see~\cite{cmbrev06}. The CMB signal is mixed with other signals
of astrophysical origin, mainly the emission from our own Galaxy and
from a large number of extragalactic objects including radio galaxies,
dusty galaxies and galaxy clusters. For the typical angular resolution
of current CMB experiments, ranging from a few arcmin to one degree,
most of these extragalactic objects appear as point sources. From the
standpoint of CMB, these point sources are contaminants that must be
removed; from the standpoint of extragalactic Astronomy, however, they
provide a valuable source of information, particularly at microwave
wavelengths where the properties of sources are poorly
understood~\cite{bluebook}. In both cases, techniques for the
detection of faint point sources are needed.

For single images taken at a given wavelength the problem is
equivalent to the general problem of detecting a number of objects,
all of them with a known waveform but unknown positions and
intensities, embedded in additive noise (not necessarily white). In
the field of microwave Astronomy, wavelet
techniques~\cite{vielva01,vielva03,MHW206,wsphere,NEWPS07}, Bayesian
approaches~\cite{hobson03,powell08}, matched
filters~\cite{tegmark98,barr03,can06} and other related linear
filtering
techniques~\cite{sanz01,naselsky02,herr02b,herr02c,can04a,can05a,can05b}
have proved to be useful. The common feature of all these techniques
is that they rely on the prior knowledge that the sources have a
distinctive spatial behaviour (i.e.~a known spatial profile, plus the
fact that they appear as compact objects as opposed to `diffuse'
random fields) that helps to distinguish them from the noise.

Most of the current CMB experiments are able to observe the sky at
several different wavelength bands. In particular, the Wilkinson
Microwave Anisotropy Probe (WMAP)~\cite{wmap0,hinshaw07} is observing
the sky at 23, 33, 41, 61 and 94 GHz and the upcoming ESA's Planck
satellite~\cite{planck_tauber05} will observe the sky in nine
frequency channels ranging from 30 to 857 GHz. Multi-wavelength data
can be used to improve the detection/separation of astrophysical
components. For example, CMB can be separated from Galactic dust and
synchrotron emission using well-established methods; a non-exhaustive
list of them include Independent Component
Analysis~\cite{fastica02,fastica07}, Maximum Entropy
Method~\cite{MEM97,vlad02,barreiro_MEM04}, Internal Linear
Combination~\cite{wmap0,eriksen04} and Wiener
filtering~\cite{bouchet99}, among others. Moreover, specific
techniques for the detection of compact sources whose spectral
behaviour is well known have been proposed in the literature; a
typical example is the detection of the so-called Sunyaev-Zel'dovich
effect due to galaxy clusters~\cite{herr02a,chema02,schaf06,melin06}.

All the previous component separation techniques, however, are in
trouble when dealing with extragalactic point sources. EPS form a very
heterogeneous population constituted by a large number of objects with
very different physical properties, from radio-emitting active
galactic nuclei to dusty star-forming galaxies. Since each source is a
unique object, with an spectral emission law that is different from the
spectral emission law of any other, the component separation problem
is underdetermined: the number $N$ of components to be separated is
much larger than the number $M$ of channels available. Attempts to
simplify the problem by grouping the extragalactic sources into
classes of objects with similar spectral behaviour are in most cases
unsatisfactory.

So far we have considered two approaches: on the one hand, it is still
possible to work channel by channel, separately, by using the
filtering techniques above mentioned. But in that case a valuable
fraction of the information that multi-wavelength experiments can
offer is wasted. On the other hand, standard multi-wavelength
component separation techniques are impracticable because the number
of physical components involved is too high. An intermediate approach
is to design filters that are able to find compact sources thanks to
their distinctive spatial behaviour while at the same time do
incorporate some multi-wavelength information, without pretending to
achieve a full component separation. In this paper we propose a new
filtering technique that makes no assumptions about the spectral
behaviour of the sources, but that makes use of some multi-wavelength
considerations, namely:
\begin{itemize}
\item When a source is found in one channel, it will be also present
  in the same position in all the other channels.
\item The spatial profile of the sources may differ from channel to
  channel, but it is a priori known. For example, for a microwave
  experiment the source profile is equal to the antenna response of
  the experiment's radiometers\footnote{In general, of its
    detectors. Some microwave experiments use bolometers instead of
    radiometers, but the distinction is irrelevant for the purposes our
    discussion.}, that is well known.
\item The second order statistics of the background in which the
  sources are embedded is well known or it can be directly estimated
  from the data by assuming that point sources are sparse.
\end{itemize}
\noindent
We find that the new technique we propose takes the form of a matrix
of filters that can be applied to any number $M$ of channels. In the
case of a single channel, the matrix of filters defaults to the
standard matched filter. In section~\ref{sec:math} we will introduce
the new formalism. In section~\ref{sec:tests} we will illustrate the
new method with two different tests: a simple toy model and a more
realistic simulation that corresponds to a small region of the sky as
it will be observed by the upcoming Planck mission. Finally, in
section~\ref{sec:conclusions} we will draw some conclusions.

\section{Matrices of filters} \label{sec:math}

\subsection{Data model}

Let us consider an experiment whose outcome is a set of $N$
two-dimensional images (channels) where there are embedded a unknown
number of point sources. For simplicity, let us consider the case of a
single point source located at the origin of the coordinates. Our data
model is
\begin{equation} \label{eq:model}
  D_k \left( \vec{x} \right) = s_k \left( \vec{x} \right) +
  n_k \left( \vec{x} \right),
\end{equation}
where the subscript $k=1,\ldots,N$ denotes the index of the
image. The term $s_k (\vec{x})$ denotes the point source,
\begin{equation} \label{eq:ps}
  s_k \left( \vec{x} \right) = A_k \tau_k \left( \vec{x} \right).
\end{equation}

In the previous equation $A_k$ is the unknown amplitude of the source
in the $k^{\mathrm{th}}$ channel and $\tau_k (\vec{x})$ is the spatial
profile of the source and satisfies the condition $\tau_k ( \vec{0} ) =
1$. We consider that the $\tau_k$ are known a priori, as is the case
in most experiments in Astronomy. For the case of point sources, each
$\tau_k$ is just the beam response (normalised to one) of the
telescope for the $k^{\mathrm{th}}$ channel.

The term $n_k (\vec{x})$ in equation (\ref{eq:model}) is the noise in
the $k^{\mathrm{th}}$ channel. In contains not only instrumental
noise, but also other astrophysical components apart from the point
sources. In the case of Microwave Astronomy, for example, $n_k
(\vec{x})$ is the combination of the CMB, Galactic emission (due to
Galactic dust, synchrotron, etc) and the instrumental noise. Since the
$N$ images correspond to the same area of the sky observed at
different wavelengths, the astrophysical components are correlated
among the different channels. Let us suppose the noise term can be
characterized by its cross-power spectrum:
\begin{equation} \label{eq:noise_ps}
\langle n_k \left( \vec{q} \right) n^{*}_l \left( \vec{q}^{~\prime}
\right) \rangle = P_{kl} \left( \vec{q} \right) \delta^2 \left( \vec{q} -
\vec{q}^{~\prime} \right),
\end{equation}
where $\mathbf{P} = (P_{kl})$ is the cross-power spectrum matrix and
$n^*$ is the complex conjugate of $n$. Besides, we assume that the
noise has zero mean,
\begin{equation}
  \langle n_k \left( \vec{x} \right) \rangle = 0.
\end{equation}

\subsection{Matrix filters} \label{sec:mfilters}

Let $\Psi_{kl} (\vec{x})$ be a set of $N \times N$ linear filters, and
let us define the set of quantities
\begin{eqnarray} \label{eq:filtered_field}
  w_k (\vec{x}) & = & \sum_l \int d\vec{x}^{~\prime} \Psi_{kl} \left( \vec{x} -
  \vec{x}^{~\prime} \right) D_l \left( \vec{x}^{\prime} \right)
  \nonumber \\
  & = & \sum_l
  \int d \vec{q} \, \, e^{-i \vec{q} \cdot \vec{x}} \, \Psi_{kl} \left(
  \vec{q} \right) D^*_l \left( \vec{q} \right).
\end{eqnarray}
The quantity $w_k$ in equation (\ref{eq:filtered_field}) is,
therefore, the sum of a set of linear filterings of the data
$D_l$. The last term of the equation is just the expression of the
filterings in Fourier space, being $\vec{q}$ the Fourier mode and
$\Psi_{lk}(\vec{q})$ and $D_l(\vec{q})$ the Fourier transforms of
$\Psi_{kl}(\vec{x})$ and $D_l(\vec{x})$, respectively.

We intend to use the combined filtered images $w_k (\vec{x})$ as
estimators of the source amplitudes $A_k$. For that purpose, we
require that the filters $\Psi_{kl}$ satisfy the condition:
\begin{equation} \label{eq:condition1}
\langle w_k (\vec{0}) \rangle =
   A_k,
\end{equation}
that is, that the $k^{\mathrm{th}}$ combined filtered image at the
position of the source is, on average over many realizations, an
\emph{unbiased} estimator of the amplitude of the source in the
$k^{\mathrm{th}}$ channel. Using equation (\ref{eq:filtered_field}),
this condition can be expressed as
\begin{equation} \label{eq:condition1b}
  A_k = \sum_l \int d\vec{q} \, \Psi_{kl} \left(\vec{q}\right) A_k
  \tau^*_l \left(\vec{q}\right).
\end{equation}
The previous condition is automatically satisfied if
\begin{equation} \label{eq:condition1c}
  \int d\vec{q} \, \Psi_{kl} \left(\vec{q}\right) \tau^*_l
  \left(\vec{q}\right) = \delta_{kl}.
\end{equation}

This condition stablishes that for a fixed image $k$ the filter
component $\Psi_{kk}$ is sensitive to the signal whereas the other
components $\Psi_{kl}$, $k\neq l$, do not affect the signal and are
sensitive only to the noise. A similar idea was proposed
by~\cite{maturi07} in the context of the extraction of the Rees-Sciama
effect where, on the the way around, the filter was defined to be
orthogonal to a noise component.

We want to estimators $w_k$ to be not only unbiased, but
\emph{efficient} as well. Then we need to minimize the variance
$\sigma_{w_k}$ of the combined filtered image. It is straightforward
to show that
\begin{equation} \label{eq:variance}
  \sigma^2_{w_k} = \sum_{l} \sum_{m} \int d\vec{q} \, \Psi_{kl}
  \left(\vec{q}\right) \Psi^*_{km} P_{lm} \left(\vec{q}\right).
\end{equation}

In order to obtain the filters that satisfy condition
(\ref{eq:condition1c}) and at the same time minimize equation
(\ref{eq:variance}) for all $k$, we solve simultaneously the
constrained minimization problem given by the Lagrangians
$\mathcal{L}_k$, $k=1,\ldots,N$:
\begin{eqnarray} \label{eq:lagrangian}
\mathcal{L}_k = \sum_{l,m} \int d\vec{q} ~\Psi^*_{km} P_{lm} \Psi_{kl}
- \nonumber \\ 2\sum_l \lambda_{kl} \left[ \int d\vec{q} ~\Psi_{kl}
  \tau^*_l - \delta_{kl} \right],
\end{eqnarray}
\noindent
where the functional dependence on $\vec{q}$ of the quantities $\Psi$,
$\tau$ and $P$ has been omitted for simplicity and where the
$\lambda_{kl}$ is a set of Lagrangian multipliers. The solution of
this problem is given by the matrix equation
\begin{equation} \label{eq:matrix_filters_eqs}
\mathbf{\Psi}^*  =  \mathbf{F} \mathbf{P}^{-1},
\end{equation}
\noindent
where
\begin{equation}
  \mathbf{F}  =  (F_{kl}),\,\,\mathbf{P} = (P_{kl}),
  \mathbf{\lambda} = (\lambda_{kl}),\,\,\mathbf{H} = (H_{kl}),
\end{equation}
\noindent
and where
\begin{eqnarray} \label{eq:matrix_filters_eqs2}
  F_{kl} & = &\lambda_{kl} \tau_l, \nonumber\\
  \mathbf{\lambda} & = & \mathbf{H}^{-1}, \nonumber \\ 
  H_{kl} & = & \int d\vec{q} \, \, \tau_k \left(\vec{q}\right) P^{-1}_{kl}
  \tau^*_l \left(\vec{q}\right) .
\end{eqnarray}
In the particular case where $P_{kl} = \delta_{kl} P_k$, that is,
where the noise is totally uncorrelated among channels, it can be seen
that the elements of the matrix of filters default to
\begin{equation} \label{eq:casematch}
  \Psi^*_{kl} \left(\vec{q}\right)
  = \delta_{kl} \frac{\tau_k \left(\vec{q}\right) / P_k
    \left(\vec{q}\right)}{
    \int d\vec{q} \, \tau^2_k \left(\vec{q}\right) / P_k
  \left(\vec{q}\right)}.
\end{equation}
In that case the matrix of filters defaults to a diagonal matrix whose
non-zero elements are the complex conjugates of the matched filters
that correspond to each channel.  In the case of circularly symmetric
source profiles and statistically homogeneous and isotropic noise, the
filters are real-valued and the whole process is equivalent to filter
each channel independently with the appropriate matched filter.  In
fact, the matched filter for the $k$-th channel is exactly given by
the right side of equation (\ref{eq:casematch}), not considering the
delta symbol. The matched filter is known to be the optimal linear
filter (that is, the one that gives maximum signal to noise
amplification) for an individual image. For more information about
matched filters in the context of point source detection in CMB
astronomy, see~\cite{tegmark98,barr03,can06} and references therein.

\subsection{Relation to other multi-wavelength approaches}
   \label{sec:MMF} 

As mentioned in the introduction, some other filtering techniques have
been proposed in the literature for the problem of detecting compact
sources in multi-wavelenght astronomical data sets. The crucial
difference is that while in all the previous approaches the spectral
behaviour of the sources must be known \emph{a priori}, here we do not
assume any prior knowledge about it. Of course, if the spectral
behaviour is perfectly known (as it is the case of the
Sunyaev-Zel'dovich effect of galaxy clusters) this knowledge can be
used to greatly improve the power of the filtering. But if there is
any uncertainty in the spectral behaviour of the sources, or if the
spectral behaviour varies from one source to another (as it is the
case of extragalactic point sources), the robustness of the method is
in great peril.
 
The so called \emph{matched
  multifilters}~\cite{herr02a,schaf06,melin06} is one of the most
commonly used multi-wavelength filtering technique. It is, in its
derivation, a close relative of the matrix filters proposed
here. Whereas for the matrix filters we have a $N\times N$ matrix of
filters that are used to produce $N$ filtered maps, in the case of
matched multifilters we have $N$ filters that are used to produce one
single, combined filtered map. The reduced dimensionality of the
matched multifilter solution is a direct consequence of the fact that
all the sources in the image are assumed to have exactly the same
spectral behaviour. The variance of this single filtered map is given
by equation (22) of~\cite{herr02a} (see that paper for a complete
description of matched multifilters).

In order to illustrate the differences between both methods, let us
imagine a very simplistic case, consisting of two images A and B that
contain white noise with identical variances $\sigma^2$. Let us assume
that the noise in A is totally uncorrelated with the noise in B and
that we place two identical sources at the same coordinates of A and
B. Images A and B can then be considered as two different `channels'
of an ideal experiment with a point source that has a perfectly flat
spectral behaviour. According to equations (\ref{eq:variance}) and
(\ref{eq:casematch}) the output of the matrix filters would be
\emph{two} images, each one filtered with the equivalent of a single
band matched filter; both images will have identical variances
$\sigma_w^2 < \sigma^2$. If, on the other hand, we use matched
multifilters (provided we know the correct spectral behaviour) it can
be seen from equation (22) of~\cite{herr02a} that the output would be
a \emph{single} map with variance $\sigma^2_{MMF}=\sigma^2_w/2$. This
simple example serves to illustrate that matched multifilters are more
powerful than matrix filters, provided the spectral behaviour is well
know\footnote{Of course, things are not that simple: in this very
example, if the spectral behaviour is really known one could always
perform a linear combination of the two uncorrelated output maps
produced by the matrix filters in order to produce a final map with
the same $\sigma^2_w/2$ variance: the key point is not the choice of
filters, but the knowledge of the spectral behaviour.}.

Conversely, if the spectral behaviour is not known then the matched
multifilters are biased (they do not recover the correct amplitude of
the sources), whereas matrix filters are unbiased independently of the
degree of knowledge of the spectral behaviour.

\subsection{Extension to the sphere}

The previous formulas have been derived for the case of flat
two-dimensional images such as the ones corresponding to a small patch
of the sky. It is straightforward to extend them to the spherical case
(i.e. the full sky), just by using harmonic instead of Fourier
transforms. Note, however, that the Galactic foregrounds vary strongly
across the sky, and therefore it may be wiser to work locally in small
areas of the sky (that can be safely projected into the tangent plane
in order to get ordinary flat images), where the spatial variation of
the foregrounds is not so drastic. This is the approach already used
in most point source detection applications in the
literature~\cite{herr02b,herr02c,can06,NEWPS07}.

\section{Numerical tests} \label{sec:tests}

In this section we will test the matrix filters in two different kind
of simulations. In both cases we will compare the performance of the
new matrix filters with the performance of the standard matched filter
applied to the individual images. The matched filter is the optimal
linear filter for individual images; therefore, any improvement with
respect to the matched filter will prove the usefulness of including
multi-wavelength information in our scheme. Other possibility would
have been to compare with wavelet filters such as the Mexican Hat
Wavelet that have been widely used in the
literature~\cite{vielva01,vielva03,MHW206,NEWPS07}. In~\cite{can06} it
has been shown, however, that wavelet filters perform very similarly
to matched filters (with a subtlety here: wavelet filters are
suboptimal in the sense of signal to noise amplification, but their
implementation is generally easier and more robust to highly
contamined images. Both effects tend to roughly compensate one with
the other, leading to an almost identical performance). Therefore, and
ir order to keep things simple, inthis work we will compare the new
technique only with standard matched filters.

The two cases we will consider are as follows. The first case will be
a toy model simulation with only two channels and very simple noise
and it will serve to illustrate the method. The second case consists
of a more realistic simulation of a small patch of the sky as will be
observed with the LFI (Low Frequency Instrument) of the upcoming
Planck mission.

\subsection{Toy model simulations} \label{sec:toy}

\begin{figure*}[!t]
  \centering
  \includegraphics[width=6.0in]{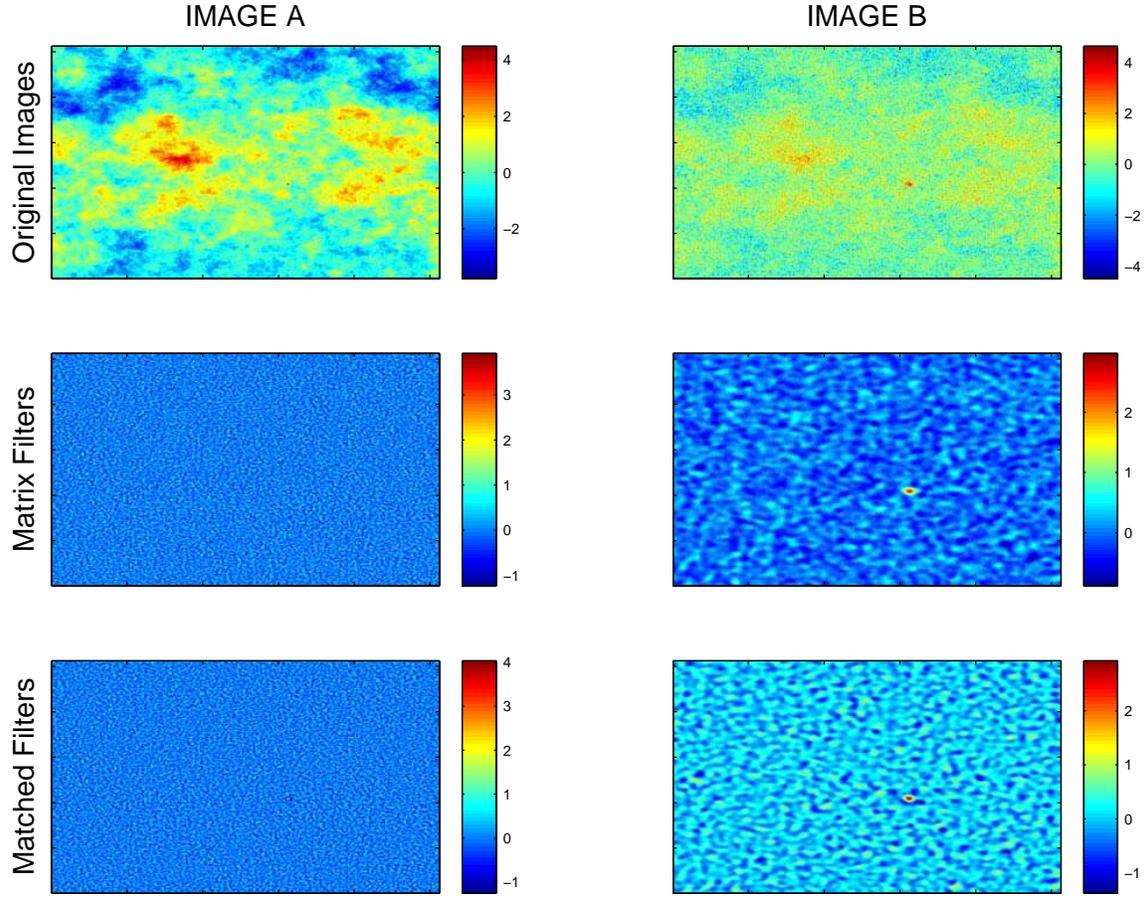}
  \caption{One of the simulations used in the toy model test. Top
  panels: images A and B before filtering. Middle panels: images A and
  B after filtering with the matrix filters. Bottom panels: images A
  and B after filtering with the matched filter that corresponds to
  each case.}
  \label{fig:simulations}
\end{figure*}

For this test we keep the things simple and simulate only a hypothetic
two-channels, ideal experiment. Each simulation consists of two
images, which we will call `image A' and `image B' respectively.  Image
A contains random color noise with power spectrum $P(q) \propto
q^{-2.5}$. For the image B we generate a random color noise with power
power spectrum $P(q) \propto q^{-0.5}$ and, since we are interested in
the case where there is some correlation between the two images, then
we add to it the first map multiplied by 0.5. For simplicity, we
normalise both images to unit variance (in arbitrary units). The
result are two correlated random noises with correlation factor
0.67. We add a Gaussian-shaped source with an intensity that is
randomly distributed in the interval $(2.75,3.75)$ (in the same
arbitrary units of the map) and $FWHM=3.33$ pixels to image A. At the
same position in image B we simulate a Gaussian-shaped source with a
different random intensity (uniformly drawn from the same interval)
and $FWHM=10$ pixels\footnote{These values have been chosen by chance:
  any other combination could serve us equally well for this
  exercise. The specific numerical results of the exercise depend on the
  sizes of the source profiles, but the qualitative behaviour of the
  method does not.}.  An example is shown in the top panels of
Fig.~\ref{fig:simulations}. We perform 250 of these simulations.

Each pair of images A and B are filtered with the matrix filters in
equation (\ref{eq:matrix_filters_eqs}). In parallel, each single image
is filtered with its corresponding matched filter (MF). Examples are
shown in the middle and bottom panels of
Fig.~\ref{fig:simulations}. Note that, in the figure, the filtered
images generated by the matrix filters and the matched filter are very
similar for image A, but quite different for image B.

\begin{figure}[!t]
  \centering
  \includegraphics[width=2.5in]{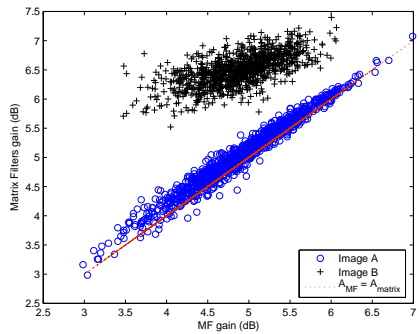}
  \caption{SIR gain of matrix filters versus the SIR gain of standard
    matched filters for the 250 simulations in our test. Blue circles
    refer to images A and black crosses refer to images B. The red
    solid line shows the amplification factor of a filter with the
    same gain factor as the matched filter.}
  \label{fig:amplif}
\end{figure}

We are interested in increasing the signal-to-interference ratio (SIR)
of the sources. Let us define the SIR gain or ``amplification'' of a
filter (or set of filters) as the ratio between the SIR of the source
after filtering and the SIR of the source before filtering. A higher
amplification factor means a better chance of detecting a source.
Fig.~\ref{fig:amplif} shows the gain factors (in dB) achieved by
matrix filters versus the gains given by standard matched filters for
images A and B. For the case of A images the ratio between the gain
obtained with the matrix filters and the gain obtained with the
matched filters is small, only 1.03 on average, whereas for the case
of B images the ratio between the two gain factors is (again on
average) 1.34, which means a significant improvement of the final SIR
of the sources. This implies that if we had fixed a given detection
threshold, the same for the two filtering techniques, with the matrix
filters we would have been able to detect sources 0.7 times less
intense than with the matched filter.

\begin{figure}[!t]
  \centering
  \includegraphics[width=2.5in]{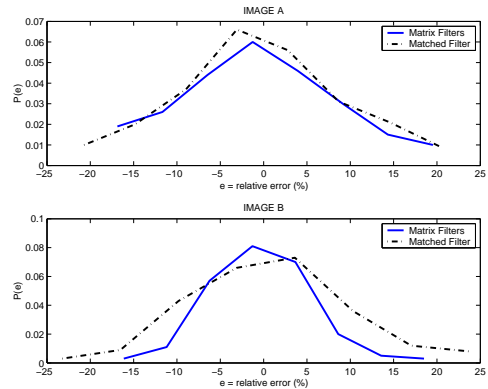}
  \caption{Relative errors in the determination of the central
  intensity of the sources for image A (top panel) and B (bottom
  panel) after filtering with the matrix filters (solid blue line) and
  the standard matched filter (dash-dotted black line).}
  \label{fig:fluxs}
\end{figure}

We are interested as well in obtaining a good estimation of the
intensity of the sources. In Fig.~\ref{fig:fluxs} the histograms of
the relative error in the estimation of the source intensities are
shown for images A and B and for matrix filters and matched
filters. The relative error is defined as $e=100 \times (I_0-I_e)/I$,
where $I_0$ is the input intensity and $I_e$ is the estimated
intensity (the central intensity after filtering). For the case of A
images the two histograms practically overlap, but for the case of B
images the histogram corresponding to matched filter spreads more,
indicating that intensity estimation with matched filters is less
reliable than with matrix filters. A more quantitative measurement of
the relative errors in each case is given by the dispersion of the
error distribution shown in Fig.~\ref{fig:fluxs}. For image A it is
$9.2\%$ and $9.6\%$ for the matrix filters and the matched filter,
respectively. For image B difference is more evident: $5.9\%$ and
$9.7\%$ for the matrix and the matched filters, respectively.

Note that this toy example is not realistic at all from the point of
view of Astronomy. For example, in a real observation of the sky the
incoming signal from astronomical objects would be filtered by the
telescope beam profile, that is not considered here. However, this
example is meant only for academic and illustrative purposes. The
effect of the telescope beam would have changed only the quantitative
results (as the input images would have been somewhat different from
the images we have used), but the qualitative behaviour of the filters
would have been the same.

\subsection{Planck simulations} \label{sec:planck}

In order to further test the applicability of matrix filters to the
detection of extragalactic objects in microwave Astronomy, we will now
use realistic simulations of the sky as it will be observed by the
ESA's Planck satellite. Again, the goal is to compare the performances
of matrix filters and the standard matched filter.

For this test, we use the ``Planck Reference Sky''
simulations\footnote{A set of state-of-the-art simulations developed
by the Planck Working Group 2 for the technical and scientific
preparation of the Planck mission.} of the Low Frequency Instrument
(LFI) Planck channels: 30, 44 and 70 GHz. The simulations include the
following astrophysical components: CMB, Galactic synchrotron,
free-free and dust emission, thermal Sunyaev-Zel'dovich effect from
galaxy clusters, radio galaxies and infrared dusty galaxies. Besides,
the simulations include instrumental noise at the levels expected for
the Planck detectors. We have selected a region of the sky centered at
the Galactic North Pole. At LFI wavelengths and at high Galactic
latitude the dominant astrophysical component is CMB, with a small
contamination from synchrotron emission. Regarding extragalactic
objects, the dominant population at this wavelengths is the one formed
by radio galaxies. For a more detailed discussion about the the Planck
Reference Sky simulations and the astrophysical components in them,
see~\cite{can06} and references therein.

\begin{figure*}[!t]
  \centering
  \includegraphics[width=6.0in]{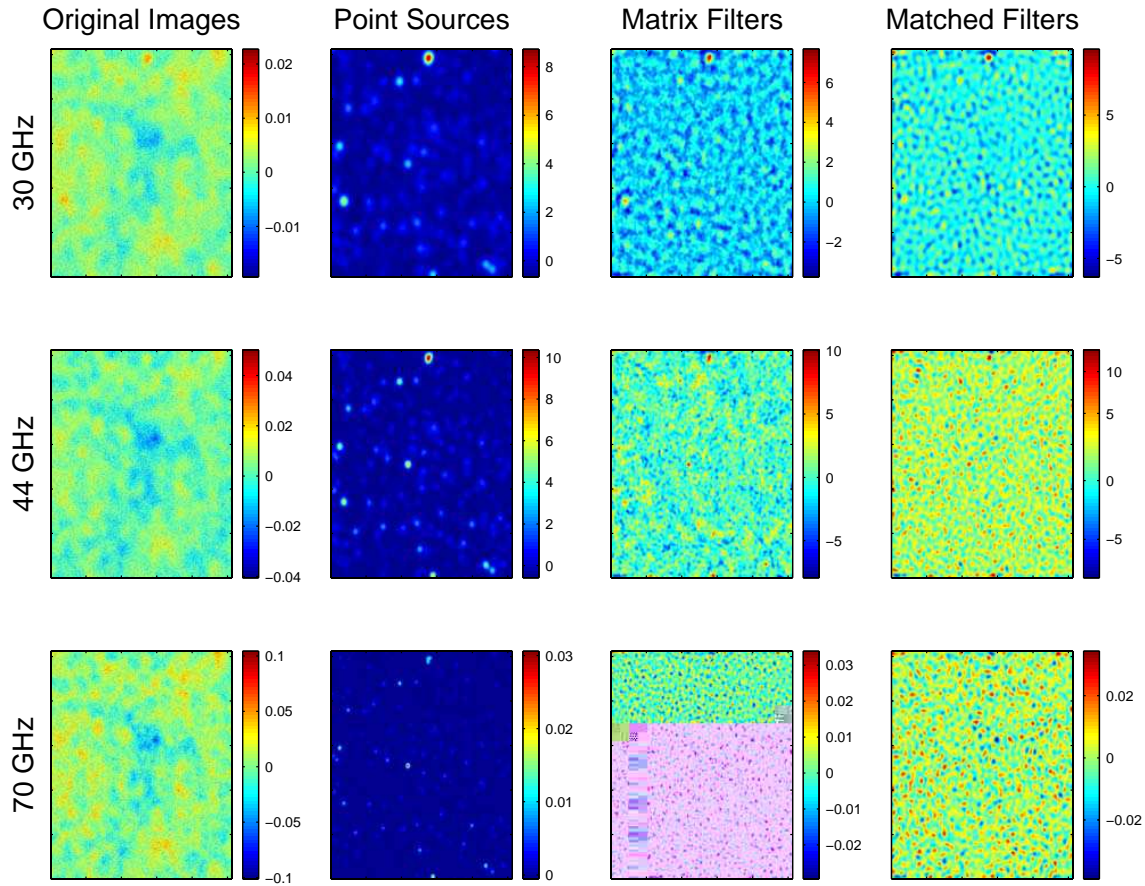}
  \caption{Planck simulations. Left panels: the original images,
  including point sources. The channels shown are, from top to bottom,
  30, 44 and 70 GHz. Second column of panels from the left: the
  point sources alone. Third column from the left: images filtered
  with the matrix filters. Fourth column from the left: images
  filtered with the matched filter.}
  \label{fig:planck}
\end{figure*}

The region of the sky we have selected covers an area of $14.66 \times
14.66$ deg$^2$ of the sky, divided into $512\times 512$ pixels of size
$1.71\times 1.71$ arcmin$^2$. The antenna FWHM for the 30, 44 and 70
GHz channels are 33, 24 and 14 arcmin, respectively. Image units are
expressed in MJy/sr\footnote{In radio astronomy, the flux unit or
  jansky (symbol Jy) is a non-SI unit of electromagnetic flux density
  equivalent to $10^{-26}$ watts per square metre per hertz. The unit
  is named after the pioneering radio astronomer Karl Jansky.}. The
images are shown in the left panels of Fig.~\ref{fig:planck}. Note
that point sources are very difficult to see in the images. For the
sake of clarity, the sources alone are shown in the second column of
panels of Fig.~\ref{fig:planck}.

\begin{figure}[!t]
  \centering
  \includegraphics[width=2.5in]{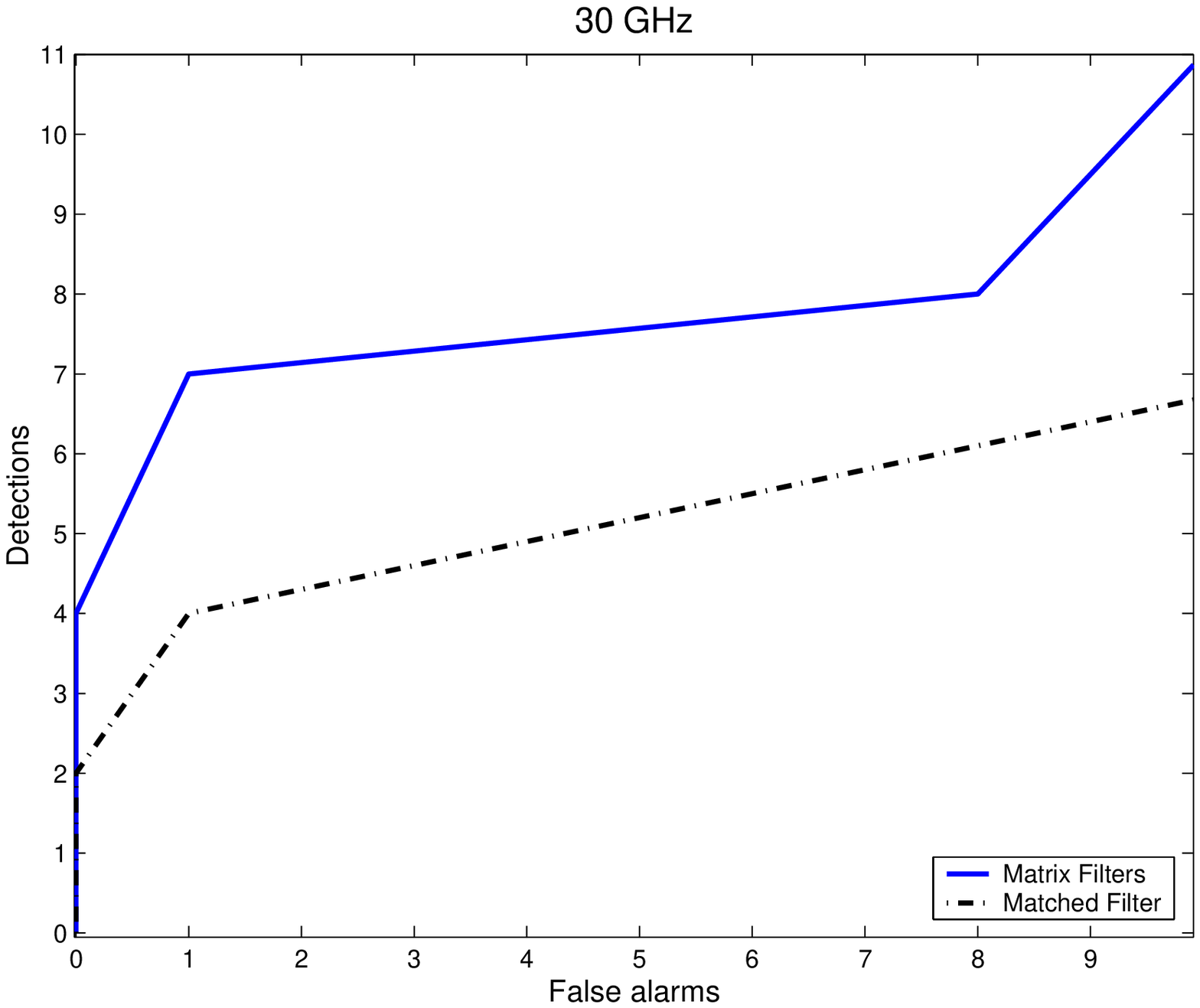}
  \caption{Number of detected sources as a function of the number of
    false alarms for the matrix filters (solid blue line) and the
    matched filter (dash-dotted black line), for the 30 GHz channel.}
  \label{fig:ROC30}
\end{figure}

\begin{figure}[!t]
  \centering
  \includegraphics[width=2.5in]{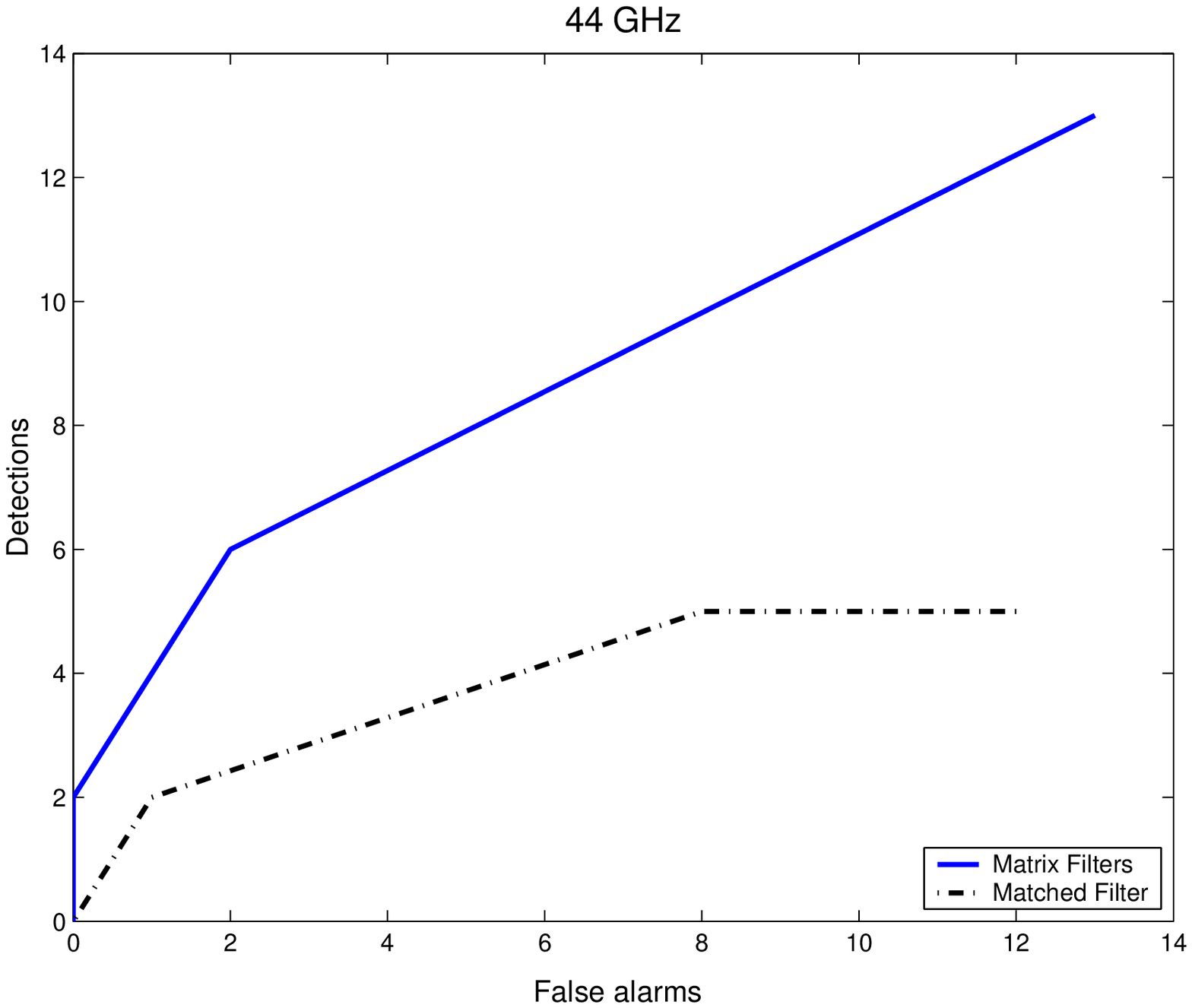}
  \caption{Number of detected sources as a function of the number of
    false alarms for the matrix filters (solid blue line) and the
    matched filter (dash-dotted black line), for the 44 GHz channel.}
  \label{fig:ROC44}
\end{figure}

\begin{figure}[!t]
  \centering
  \includegraphics[width=2.5in]{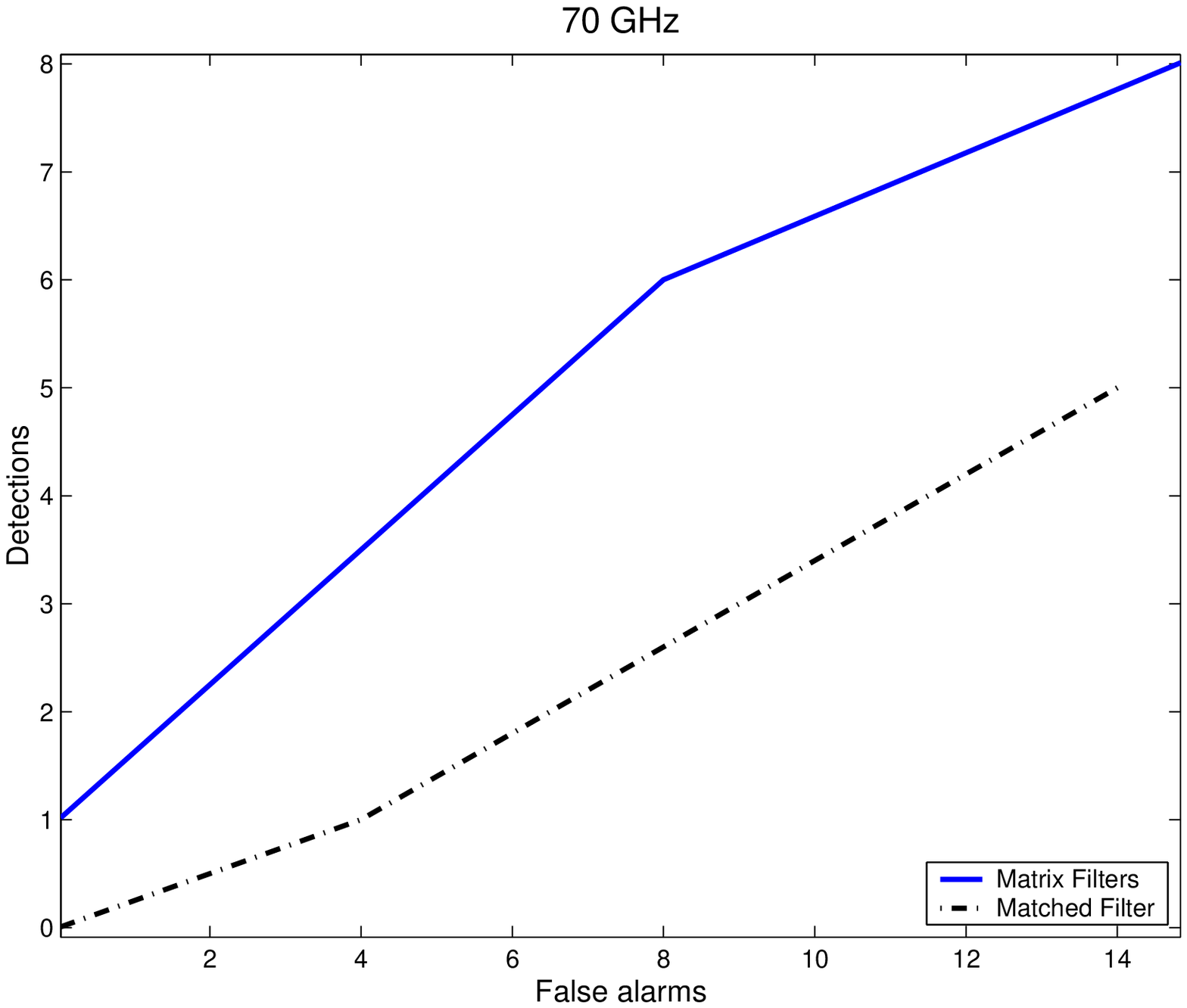}
  \caption{Number of detected sources as a function of the number of
    false alarms for the matrix filters (solid blue line) and the
    matched filter (dash-dotted black line), for the 70 GHz channel.}
  \label{fig:ROC70}
\end{figure}

We have filtered the images with the matrix filters and with the
standard matched filter. The filtered images are shown in the third
and fourth column of Fig.~\ref{fig:planck}, respectively. For
different detection thresholds and for both filters we have calculated
the number of detected sources and the number of false alarms. The
results are shown in Figs.~\ref{fig:ROC30},~\ref{fig:ROC44}
and~\ref{fig:ROC70}.  For the three channels, matrix filters
consistently give a higher number of true detections for a fixed
number of false alarms. For all the considered cases, the positive
detection rate over the minimum considered detection threshold is
equal to one, that is: we detect all the sources we put in the
simulation.

Another indicator of quality is the completeness flux, that is, the
threshold above which all the input sources are detected. For the
matrix filters, completeness fluxes are 0.16, 0.19 and 0.58 Jy for the
30, 44 and 70 GHz channels, respectively. For the matched filter,
completeness fluxes are 0.32, 0.40 and 0.58 Jy for the same
channels. This means that in two of the three channels considered
matrix filters are able to detect fainter point sources than the
matched filter. In the 70 GHz channel matrix filters reaches the same
flux limit than the matched filter, but from Fig.~\ref{fig:ROC70} it
is clear that for a fixed number of false alarms they lead to a higher
number of positive detections.

\section{Conclusion} \label{sec:conclusions}

In this work we have introduced a new filtering technique, the matrix
filters, that can be of utility for the detection of extragalactic
point sources in multi-wavelength experiments of the Cosmic Microwave
Background. Matrix filters are designed to maximize the
signal-to-interference ratio of compact sources embedded in a set of
images (``channels'') by taking into account the cross-correlations
between the different channels. For the case of a single-channel
experiment and circularly symmetric sources, the matrix of filters
defaults to the standard matched filter. For the case of $M$ totally
uncorrelated channels and circularly symmetric sources, the matrix of
filters becomes a diagonal matrix whose non-zero elements are the
matched filters corresponding to each channel. In a general case, the
non diagonal elements of the matrix are non zero.  Since the matrix
filters contain as a particular case the matched filter and they are
designed to maximize the signal-to-interference ratio, then the
individual amplifications for each channel (defining by amplification
the quotient between the signal-to-interference ratios after and
before filtering) are greater or (in the worst case) equal to the
amplifications obtained by the standard matched filters.

We have tested the matrix filters in two simulated cases: a simplistic
two-channel case with ideal noise and a more realistic simulation of
the sky as it will be observed by the LFI instrument of the upcoming
ESA's Planck mission. In the first test we observe that matrix filters
clearly outperform the standard matched filter for one of the
channels, while for the other channel their performance is very
similar to that of the matched filter. Other simulation tests with
different parameters we have carried out show a similar behaviour:
there is a significant gain for at least one of the channels with
respect to standard matched filters.

For the second test case we have considered one single patch of the
sky, observed at 30, 44 and 70 GHz. The purpose of this test was just
to give an example of how the matrix filters work in a more realistic
case. In this example, matrix filters outperform the matched filter
both in the flux limit they can reach and in the ratio between false
alarms and true detections. This result seems to indicate that matrix
filters could help to obtain better catalogues of extragalactic point
sources in future CMB experiments. However, this result has been
obtained for a single simulation of a small portion of the sky, and
therefore it may not be extrapolable to the whole sky or to any other
experiment. The study of the application of matrix filters to the
whole microwave sky for the Planck mission is the subject of a future
work.

\section*{Acknowledgment}

The authors acknowledge partial financial support from the Spanish
Ministry of Education (MEC) under project ESP2004-07067-C03-01 and
from the joint CNR-CSIC research project 2006-IT-0037. JLS, on
sabbatical leave, acknowledges financial support from the Spanish
Ministry of Education and thanks the CNR ISTI for the hospitality
during the sabbatical. We would like to thank J.~Gonz\'alez-Nuevo and
M.~L\'opez-Caniego for useful comments.

\ifCLASSOPTIONcaptionsoff
  \newpage
\fi

\bibliographystyle{IEEEtran}
\bibliography{IEEEabrv,herranz_bib}
%\begin{thebibliography}{32}
%
%\end{thebibliography}

% biography section
%
% If you have an EPS/PDF photo (graphicx package needed) extra braces are
% needed around the contents of the optional argument to biography to prevent
% the LaTeX parser from getting confused when it sees the complicated
% \includegraphics command within an optional argument. (You could create
% your own custom macro containing the \includegraphics command to make things
% simpler here.)
%\begin{biography}[{\includegraphics[width=1in,height=1.25in,clip,keepaspectratio]{mshell}}]{Michael Shell}
% or if you just want to reserve a space for a photo:

\begin{IEEEbiography}{Diego Herranz}
received the B.S. degree in 1995 and the M.S. degree in physics from
the Universidad Complutense de Madrid, Madrid, Spain, in 1995 and the
Ph.D. degree in astrophysics from Universidad de Cantabria, Santander,
Spain, in 2002. He was a CMBNET Postdoctoral Fellow at the Istituto di
Scienza e Tecnologie dell'Informazione ``A. Faedo'' (CNR), Pisa,
Italy, from 2002 to 2004. He is currently at the Instituto de F\'\i
sica de Cantabria, Santander, Spain, as UC Teaching Assistant.  His
research interests are in the areas of Cosmic Microwave Background
astronomy and extragalactic point source statistics as well as the
application of statistical signal processing to astronomical data,
including blind source separation, linear and nonlinear data
filtering, and statistical modeling of heavy-tailed processes.
\end{IEEEbiography}

% if you will not have a photo at all:
\begin{IEEEbiography}{Jos\'e Luis Sanz}
received the Ph.D. degree in theoretical physics from Universidad
Autonoma de Madrid, Spain, in 1976. He was a M.E.C. Postdoctoral
Fellow at the Queen Mary College), London, U.K., during 1978. He is
currently at the Instituto de F\'\i sica de Cantabria, Santander,
Spain, as UC Professor on Astronomy since 1987.  His research
interests are in the areas of Cosmic Microwave Background astronomy
(anisotropies, non-Gaussianity), extragalactic point sources and
clusters of galaxies (blind/non-blind detection, estimation,
statistics) as well as the developement of techniques in signal
proccesing (wavelet design, linear/non-linear filters, time-frequency,
sparse representations) and application of such tools to astronomical
data.
\end{IEEEbiography}

\end{document}